\documentclass[english,aps,prl,reprint,nofootinbib,nobibnotes,notitlepage,preprintnumbers,showpacs]{revtex4-1}
\usepackage{amsmath, amssymb, amsthm, braket}
\usepackage{graphicx}
\usepackage{listings}
\usepackage[T1]{fontenc}

\usepackage[unicode=true,bookmarks=true,
bookmarksnumbered=false,bookmarksopen=false,
breaklinks=false,
pdfborder={0 0 1},
backref=false,colorlinks=true]{hyperref}
\hypersetup{pdftitle={Impact of Natural Planck Scale Cutoffs that are Fully Covariant on Inflation}, pdfauthor={Aidan Chatwin-Davies, Achim Kempf, and Robert T. W. Martin}, citecolor=blue,linkcolor=blue,urlcolor=blue,citecolor=blue}

\usepackage[usenames,dvipsnames]{color}

\newcommand{\mbf}[1]{\mathbf{#1}}
\newcommand{\be}{\begin{equation}}
\newcommand{\ee}{\end{equation}}

\newcommand{\Eq}[1]{Eq.~\eqref{#1}}
\newcommand{\Fig}[1]{Fig.~\ref{#1}}

\begin{document}

\preprint{\hbox{CALT-TH-2016-016}}

\title{Natural Covariant Planck Scale Cutoffs and the Cosmic Microwave\\Background Spectrum }
\date{May 1, 2017}
\author{Aidan Chatwin-Davies}
\affiliation{Walter Burke Institute for Theoretical Physics,\\
\mbox{California Institute of Technology, Pasadena, California 91125, USA}}
\author{Achim Kempf}
\affiliation{Departments of Applied Mathematics and Physics, University of Waterloo, Waterloo, Ontario N2L 3G1, Canada
}
\author{Robert T. W. Martin}
\affiliation{Department of Mathematics and Applied Mathematics, University of Cape Town, Cape Town, Western Cape 7701, South Africa}
\thanks{
\href{mailto:achatwin@caltech.edu}{\tt achatwin@caltech.edu}, \href{mailto:akempf@perimeterinstitute.ca}{\tt akempf@perimeterinstitute.ca},\\ \href{mailto:rtwmartin@gmail.com}{\tt rtwmartin@gmail.com}
}

\begin{abstract}
We calculate the impact of quantum gravity--motivated ultraviolet cutoffs on inflationary predictions for the cosmic microwave background spectrum.
We model the ultraviolet cutoffs fully covariantly to avoid possible artifacts of covariance breaking. Imposing these covariant cutoffs results in the production of small, characteristically $k-$dependent oscillations in the spectrum.
The size of the effect scales linearly with the ratio of the Planck to Hubble lengths during inflation.
Consequently, the relative size of the effect could be as large as one part in $10^5$; i.e., eventual observability may not be ruled out. 
\end{abstract}
\maketitle

It is widely expected that the very notion of distance in space and time breaks down at or before the Planck scale, due to quantum fluctuations of the metric (see \cite{Hossenfelder2013} for a review).
To understand the structure of spacetime at the Planck scale will, therefore, require a theory of quantum gravity.
There are several current approaches to a consistent theory of quantum gravity, including string theory, loop quantum gravity, and others \cite{Rovelli1997,Carlip2015}.
It has proven exceedingly difficult, however, to test models for Planck-scale physics experimentally, chiefly due to the extremely small scales involved. 

One of the most promising approaches to experimentally probing quantum gravity theories is to look for small imprints  that Planck-scale physics may have left in the cosmic microwave background (CMB) and the subsequent structure formation \cite{Jacobson1999,Brandenberger2001a,Kempf2001a,Martin2001,Kempf2001,Shiu2005}.
This is because, according to the standard inflationary scenario, the quantum fluctuations that seeded the CMB's inhomogeneities likely originated only about 5 or 6 orders of magnitude away from the Planck scale \cite{Shiu2005}, namely when the comoving modes' fluctuations froze at their Hubble horizon crossing. 

The magnitude of the imprint of Planck-scale physics in the CMB has been estimated using various approaches. These generally model the influence that Planck-scale physics exerts on inflationary quantum field theory through an ultraviolet (UV) cutoff at the Planck scale and through generalized dispersion relations \cite{Padmanabhan1988, Padmanabhan1989, Brandenberger2001, Brandenberger2002, Easther2001, Kempf2001, Niemeyer2001, Easther2002, Easther2003, Brandenberger2005, Easther2005, Greene2005, Padmanabhan2005}. Of crucial importance is the question of how the magnitude of the effect scales with the ratio of the Planck and Hubble lengths during inflation \cite{Danielsson2002,Shiu2005}.
Let us denote the Hubble scale at the end of inflation by $L_\text{Hubble}$ and define $\sigma = L_{\text{Planck}}/L_{\text{Hubble}}$.
Several studies found that the imprint of quantum gravity in the CMB should be of the order of $\sigma^\alpha$ with either $\alpha=1$ \cite{Easther2001} or $\alpha =2$ \cite{Kempf2001}.
If indeed $\alpha=1$, then we may be a mere five orders of magnitude away from measuring Planck-scale physics, which is relatively close when compared to accelerator physics, where there are about 15 orders of magnitude to cross.  

Models for how Planck-scale physics influences inflationary quantum field theory, such as modified dispersion relations, generally break local Lorentz invariance \cite{Shiu2005}.
This makes it unclear to what extent the predicted imprints on the CMB are due to Planck-scale physics and to what extent the predicted imprints are caused by the breaking of local Lorentz invariance.
Therefore, to isolate the effect of Planck-scale physics on the CMB, we employ functional analytic methods that allow us to model natural UV cutoffs fully covariantly. 

Our main results are that covariant UV cutoffs can produce small characteristic oscillations in the fluctuation spectrum and that this imprint on the CMB is of first order, i.e., $\alpha=1$. The overall amplitude is sensitive to the timing of the comoving modes' quantum-to-classical transition. 
This transition is generally expected to have occurred 
soon after horizon crossing \cite{Polarski1995,Kiefer1998,Kiefer2008,Nelson2016,Boddy2016}. This is because, after horizon crossing, comoving modes quickly become extremely squeezed, which makes them extremely susceptible to environment-induced decoherence \cite{Joos2003}. We will be able to conclude, therefore, that the relative size of the imprint of a covariant Planck-scale cutoff in the CMB could be as large as one part in $10^5$ so that eventual observability may not be ruled out.

\emph{Fully covariant natural ultraviolet cutoffs.} 
Our aim is to covariantly model possible first order corrections to standard quantum field theory (QFT) as the Planck scale is approached from low energies.
To this end, recall that in the path integral formulation of QFT, the path integral is normally assumed to run over all field configurations.
This includes on-shell fields, i.e., fields that extremize the action, as well as fields that are arbitrarily far off-shell.
In order to ensure the preservation of covariance, 
we will here consider UV cutoffs that remove or suppress field configurations---and therefore field fluctuations---that are off-shell by an amount that is on the order of or past the Planck scale.
Technically, the eigenfunctions of the d'Alembertian to eigenvalues beyond the Planck scale will be considered too far off-shell and will therefore be modeled as being suppressed or eliminated entirely from the path integration. 
Such cutoffs are manifestly covariant and diffeomorphism invariant since the spectra of covariant differential operators such as the d'Alembertian, and the corresponding operators for fields other than scalars, are independent of the choice of coordinates.

Concretely, let $\Box$ denote a self-adjoint d'Alembertian on a Lorentzian manifold, $\mathcal{M}$. 
Its eigenfunctions form an orthonormal basis for $L^2(\mathcal{M})$ and can be taken to span the space of functions in the quantum field theoretic path integral.
A cutoff on the spectrum of $\Box$ is operationally defined via (real linear combinations of) spectral projectors, 
\begin{equation} \label{eq:projectors}
f (\Box ) = \sum_{\lambda \in \mathrm{spec} (\Box)} f(\lambda) \,  \langle \psi_\lambda , \cdot \, \rangle \, \psi_\lambda \, ,
\end{equation}
where $f$ is the non-negative function which defines the cutoff, $\psi_\lambda$ is an eigenfunction of $\Box$, and $\langle \cdot , \cdot \rangle$ denotes the inner product on $L^2(\mathcal{M})$.

Choosing $f(\lambda) = \theta(\Omega^2 - |\lambda|)$ produces a sharp cutoff on the spectrum of $\Box$ at $|\lambda| = \Omega^2$.
We will focus on this most extreme case of an UV cutoff.
It is straightforward to smooth out the step function so that the spectral cutoff is not as sharp and the positive operator  $f (\Box )$  describes models of smoother UV cutoffs.  
These covariant cutoffs amount to suppressing far off-shell field fluctuations from the path integral. 
\footnote{We remark that, since proton decay is mediated by far off-shell processes, such a covariant ultraviolet cutoff may help explain the exceedingly large proton lifetime.}

The class of cutoffs that we consider is the most general type of kinematic covariant cutoff within the framework of QFT.
We are not going beyond the framework of QFT because inflation appears to have happened well within the range of validity of QFT.
Among these cutoffs, we focus on the most extreme case, the sharp cutoff, to obtain a prediction for the maximal impact on the fluctuation spectrum. 

As an aside, we also note that with the above choice of a sharp cutoff $f$, the quantity $\Omega$ can be interpreted as a covariant bandlimit: A conventional bandlimit, i.e., a minimum wavelength imposed on functions in $\mathbb{R}^n$, can be thought of as a cutoff on the spectrum of the Laplacian, $\triangle$, on $\mathbb{R}^n$.
The eigenfunctions of $\triangle$ are the plane waves $\exp(i \mbf k \cdot \mbf x)$ and cutting off the spectrum of $\triangle$ is to impose a limit on the length of the wave vector $k^2$.
As a consequence, Shannon's sampling theorem applies: Any function that is $\Omega$-bandlimited is determined everywhere if known on any regular discrete lattice $\{x_n\}$ with a spacing smaller than $ 1/(2\Omega)$ \cite{Shannon1949,Landau1967,Jerri1977}, in each direction. 
The d'Alembertian generalizes the Laplacian to Lorentzian spacetimes, leading to a covariant generalization of sampling theory \cite{Kempf2013,Kempf2008info}: 
Each mode of a particular spatial wavelength possesses a corresponding bandwidth and therefore obeys a sampling theorem in time. Those modes whose wavelengths are smaller than the Planck length possess an exceedingly small bandwidth, which effectively freezes them out. 
This ultraviolet behavior is beautifully covariant, as the notions of spatial wavelength and temporal bandwidth Lorentz transform appropriately.  

Returning to our main program, we express the space of covariantly-bandlimited scalar fields on $\mathcal{M}$ in terms of eigenfunctions of the d'Alembertian as
\begin{equation}
B_\mathcal{M}(\Omega) = \mathrm{span}\left\{\psi \; | \; \Box \psi = \lambda \psi, \lambda \in [-\Omega^2,\Omega^2] \right\} .
\end{equation}
Here, $\Omega$ sets the ultraviolet scale.
In general, the spectrum of a self-adjoint d'Alembertian is not bounded below for Lorentzian-signature manifolds, and so the spectrum must be cut off from above and below.

In the path integral formulation of QFT, we implement the covariant bandlimitation by only integrating over covariantly-bandlimited fields instead of all field configurations.
For example, the covariantly-bandlimited Feynman propagator of a quantized scalar field, which we denote by $G_F^\Omega$, is given by
\begin{equation}\label{eq:pathint}
iG_F^\Omega(x,x^\prime) =  \frac{\int_{B_\mathcal{M}(\Omega)} \mathcal{D}\phi ~ \phi(x) \phi(x^\prime) e^{iS[\phi]}}{\int_{B_\mathcal{M}(\Omega)} \mathcal{D}\phi ~ e^{iS[\phi]}} \, .
\end{equation}
Covariant bandlimitation here amounts to excluding the most extreme off-shell fluctuations from the quantum field theoretic path integral.
Concretely, $G_F^\Omega$ can be computed by acting on the conventional propagator $G_F$ to the left and right with the spectral projectors $\theta (\Omega ^2 - \Box ) \equiv P_\Omega$, where $G_F(x,x^\prime)$ is understood to be the kernel of an integral operator.

\emph{Application to inflation.}
Let $\mathcal{M}$ be an inflating Friedmann-Robertson-Walker (FRW) spacetime with the line element $ds^2 = a^2(\eta) [-d\eta^2 + d\mbf{x}^2 ]$, where the conformal time $\eta$ takes values in an interval $\mathcal{I} \subseteq (-\infty,0)$. 
Consider a massless scalar field $\phi$ on this background.
Such a field is a proxy for quantities such as the Mukhanov-Sasaki variable, which describes combined quantized perturbations of the inflaton and scalar metric degrees of freedom, or tensor perturbations of the metric, which describe primordial gravitational waves.

The strength of the field's quantum fluctuations is quantified by its fluctuation spectrum:
\begin{equation}
\begin{array}{rl}
\delta\phi_k(\eta) &= \frac{1}{2\pi}k^{3/2}|v_k(\eta)| \\[2mm]
&= \sqrt{4\pi} k^{3/2} |G_F(\eta=\eta^\prime;k)|^{1/2} \, .
\end{array}
\end{equation}
Here, $v_k(\eta)$ is the field's mode function and $k$ is comoving wavelength.
We define the covariantly-bandlimited fluctuation spectrum by replacing $G_F$ in the equation above with $G_F ^\Omega = P_\Omega G_F P_\Omega$.
Note that the spectrum of the d'Alembertian in a FRW spacetime is preserved (up to degeneracy) under spatial Fourier transforms, i.e., if $\Box u(\eta,\mbf{x}) = \lambda u(\eta,\mbf{x})$, then $\Box_k u_\mbf{k}(\eta) = \lambda u_\mbf{k}(\eta)$.
This allows us to calculate the bandlimited propagator comoving mode by comoving mode.

We now consider the flat slicing of de Sitter spacetime in 1+3 dimensions, which is the most computationally tractable model for an inflating FRW spacetime.
The scale factor is $a(\eta) = - 1/H\eta$ with $-\infty < \eta < 0$, and $H$ is the Hubble parameter.
We take the state of the field to be the Bunch-Davies vacuum \cite{Birrell1982} so that the Feynman propagator without cutoff reads, as usual:
\begin{widetext}
\begin{equation} \label{eq:dSGF}
G_F(\eta,\eta^\prime;k) = -\frac{i\pi}{4} \frac{\sqrt{\eta\eta^\prime}}{a(\eta)a(\eta^\prime)} \left[\theta(\eta-\eta^\prime)H_{3/2}^{(1)}(k|\eta|)H_{3/2}^{(2)}(k|\eta^\prime|) + \theta(\eta^\prime-\eta)H_{3/2}^{(2)}(k|\eta|)H_{3/2}^{(1)}(k|\eta^\prime|) \right].
\end{equation}
\end{widetext}
Here, $H_{3/2}^{(1)}$ and $H_{3/2}^{(2)}$ denote Hankel functions of the first and second kind, respectively.

Our strategy is as follows: For each comoving mode $k$, construct the spectral projectors $P_\Omega$ from the eigenfunctions and eigenvalues of $\Box_k$, and then apply these to the left and right of $G_F(\eta,\eta^\prime; k)$ to obtain $G_F^\Omega$.
Equivalently, we can write $P_\Omega = I - P_\Omega ^\perp$, where $P_\Omega ^\perp$ projects onto eigenspaces corresponding to $|\lambda| > \Omega^2$, so that
\begin{equation} \label{eq:PGP}
G_F^\Omega = G_F - (P_\Omega ^\perp  G_F + G_F  P_\Omega ^\perp -  P_\Omega ^\perp G_F  P_\Omega ^\perp )
\end{equation}
and the quantity in brackets gives the correction to the full propagator.

We notice first that each $k$-d'Alembertian on $L^2((-\infty,0), a^4(\eta)d\eta)$ is not uniquely self-adjoint. 
In functional analytic language \cite{Naimark1968,Sturm1,Sturm2}, the minimal symmetric operator generated by each $\Box_k$ has deficiency indices $(1,1)$, which implies the existence of a one-parameter family of self-adjoint extensions of $\Box_k$, each corresponding to a generalized boundary condition.
We identify the correct self-adjoint extension by matching the generalized boundary condition to the boundary condition implied by $G_F$ as given in \Eq{eq:dSGF}. The fact that $G_F$ is a right inverse of the d'Alembertian, i.e., that $\Box_k G_F(\eta,\eta^\prime; k) = a^{-4}(\eta) \delta(\eta-\eta^\prime)$, means that $G_F$ is diagonal in the same basis as $\Box_k$ and so it shares the same boundary condition.

For each $\lambda \in \mathbb{R}$, the eigenfunction equation $\Box_k u(\eta) = \lambda u(\eta)$  yields a Sturm-Liouville differential equation,
\begin{equation} \label{eq:SL}
(a^2 u^\prime)^\prime + k^2 a^2 u + \lambda a^4 u = 0. \ 
\end{equation}
This can be solved to obtain two linearly independent solutions \cite{Naimark1968}.
For $\lambda < 9H^2/4$, \Eq{eq:SL} admits one normalizable solution and so the self-adjoint extensions of $\Box_k$ will have point spectrum in this range.
For $\lambda \geq 9H^2/4$, both solutions are non-normalizable, and so this range can only contain continuous spectrum.
The possible normalizable solutions for $\lambda < 9H^2/4$ are
\begin{equation}
\psi_n(\eta) = H^2 \sqrt{2p_n} \, |\eta|^{3/2} J_{p_n}(k|\eta|) \, ,
\end{equation}
where $p_n = p_0 + 2n$ with $n \in \mathbb{N}$ ensures orthonormality and the value of $p_0 \in (0,2]$ fixes the self-adjoint extension. (Here, $J_p$ denotes the Bessel-$J$ function of order $p$). The corresponding eigenvalues are $\lambda_n = H^2(\tfrac{9}{4}-p_n^2)$.
We determined the correct choice of self-adjoint extension by examining the action of $G_F^h - \lambda_n^{-1} I$ on test eigenfunctions as a function of $p_0$, where $G_F^h$ denotes the Hermitian part of $G_F$.
When $p_0$ takes the value that is implied by $G_F$ as given in \Eq{eq:dSGF}, then $(G_F^h-\lambda_n^{-1}I)\psi_n(\eta \, ; p_0)$ must be in the kernel of $\Box_k$.
By varying $p_0$ and checking when this last condition is satisfied, we found that $p_0 = 3/2$ (so that $\lambda _0 =0$) is the self-adjoint extension that is implied by \Eq{eq:dSGF}.

Orthonormality then implies that all $\lambda \geq 9H^2/4$ are in the continuous spectrum, and that the corresponding eigenfunctions are
\begin{align}
\nonumber \psi_{q(\lambda)}(\eta) = & H^2 \sqrt{\tfrac{1}{2} q \tanh(\pi q)} \left[\mathrm{sech}(\tfrac{\pi}{2} q) |\eta|^{3/2}\mathrm{Re}\,J_{iq}(k|\eta|) \right. \\
& \left.- \mathrm{csch}(\tfrac{\pi}{2} q) |\eta|^{3/2} \mathrm{Im}\,J_{iq}(k|\eta|) \right],
\end{align}
with $q(\lambda) = (\tfrac{\lambda}{H^2}-\tfrac{9}{4})^{1/2}$.
The eigenfunctions of the continuous spectrum have been normalized so that
\begin{equation}
\int_{-\infty}^0 \psi_q(\eta) \psi_{q^\prime}(\eta)~a^4(\eta)d\eta = \delta(q-q^\prime) \, .
\end{equation}
We deduced this normalization numerically by requiring that $\int_{q-\epsilon}^{q+\epsilon} \int_{-\infty}^0 \psi_q(\eta) \psi_{q^\prime}(\eta)~a^4(\eta)d\eta = 1$.
Upon exchanging the order of integration, the double integral becomes Riemann-integrable and thus calculable numerically.

\begin{figure}
\includegraphics[width=\columnwidth]{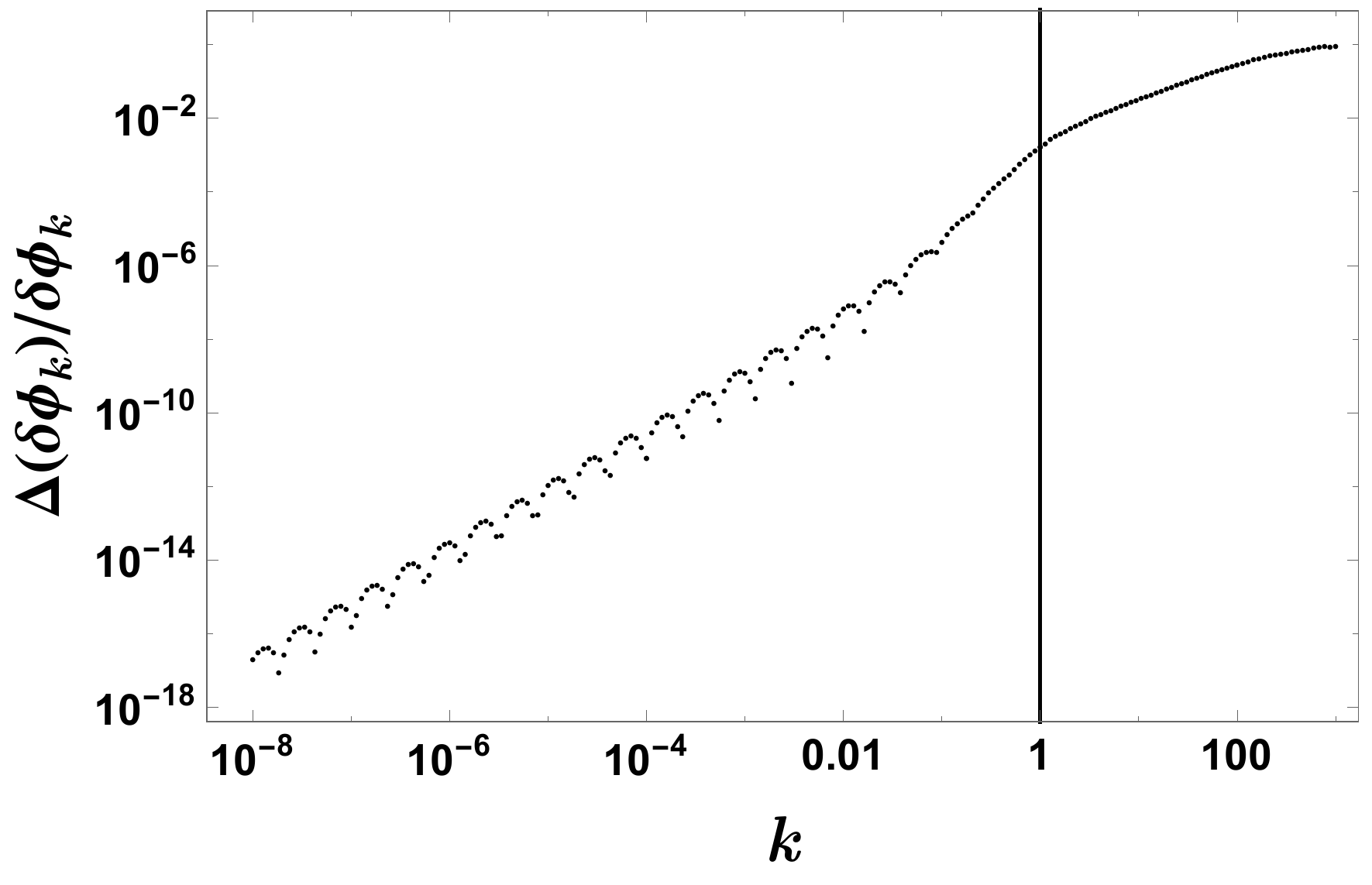}
\caption{Relative change in $\delta \phi_k$ as a function of comoving mode number with $|\eta|=1$ fixed. Horizon crossing ($k|\eta|=1$) is marked with a solid line. The ratio of Planck to Hubble scales is set to $H/\Omega = 10^{-2}$ so that the numerical computation remains tractable. The oscillations are due to an oscillatory integrand that appears in $G_F^\Omega$ which dominates at small $k$.
Roughly, the oscillations may be understood as an interference effect, with dips occurring when an integer number of wavelengths fit between the Planck and Hubble scales.}
\label{fig:reldiff}
\end{figure}

Assembling the results, the projectors $P^\perp_\Omega$  are given by
\begin{equation} \label{eq:projectorsconcrete}
P^\perp_\Omega(\eta,\eta^\prime) = \sum_{n>N} \psi_n(\eta) \psi_n(\eta^\prime) + \int_Q^{\infty} \psi_q(\eta)\psi_q(\eta^\prime)~dq \, ,
\end{equation}
where $Q = q(\Omega^2)$ and $N = \max\{n : |\lambda_n| < \Omega^2 \}$.
The full correction to the propagator in \Eq{eq:PGP} can then be calculated using a combination of exact antiderivatives when possible and numerical integration otherwise.
The contribution from the point spectrum is several tens of orders of magnitude smaller than the contribution from the continuous spectrum, and is therefore negligible in \Eq{eq:projectorsconcrete}.
A plot of $\Delta(\delta\phi_k)/\delta\phi_k$ is shown in \Fig{fig:reldiff}, where
\begin{align}
\nonumber \Delta(\delta\phi_k(\eta)) = & \sqrt{4\pi} k^{3/2} \left(|G_F^\Omega(\eta=\eta^\prime;k)|^{1/2} \right. \\[-1mm]
& \left. - |G_F(\eta=\eta^\prime;k)|^{1/2} \right).
\end{align}
The quantity $\Delta(\delta\phi_k)/\delta\phi_k$ characterizes the magnitude of the impact of the covariant UV cutoff on inflationary predictions for the CMB.

\Fig{fig:reldiff} shows that the impact of the covariant bandlimit on inflationary perturbations is sensitive to when the perturbations were in effect measured and became classical.
As is well known, after horizon crossing, comoving modes quickly become highly squeezed.
This squeezing makes the quantum statistics of field fluctuations indistinguishable from those of a classical stochastic ensemble, which leads to an ``apparent'' quantum-to-classical transition \cite{Kiefer2008}. 
If the comoving modes had continued to evolve as a closed system until re-heating, then our calculations, which hold as long as the modes are independent closed systems, would be valid up until re-heating.
At this point the re-heating interactions will decohere the inflationary perturbations, which would then be described according to open system dynamics.
In this scenario, the imprint that Planck-scale physics could have left in the CMB would, in effect, be measured and fixed as late as re-heating.
The imprint would therefore be exponentially suppressed. 

However, the modes' squeezing at horizon crossing is generally expected to have also led to a quantum-to-classical transition through environmental decoherence already shortly after horizon crossing, therefore requiring an open system description at that stage.
For our calculation, this means that the imprint of Planck-scale physics in the CMB was in effect measured soon after horizon crossing.
We therefore predict that this imprint is not suppressed.

For completeness, let us briefly discuss why decoherence is expected to have occurred soon after horizon crossing (for a more extensive review, see \cite{Kiefer2008}).
During inflation, there are several environmental sources of decoherence for the comoving modes.
These include nonlinear gravitational interactions among the modes as well as interactions with the fields of other species.
(There is also the effect of decoherence from tracing over degrees of freedom beyond the cosmological horizon.)
The weakness of these environmental interactions is offset by the well-known \cite{Joos2003} extreme sensitivity of highly-squeezed states, such as the comoving modes after horizon crossing, to environmental decoherence.
We remark that these decoherence mechanisms are indeed such that the pointer basis is approximately the field eigenbasis and that the modes' standing wave behavior can account for the acoustic oscillations in the CMB, as required by phenomenology \cite{Kiefer2008}.

Within this standard scenario for the quantum-to-classical transition in inflation, we can then conclude that the cutoff-induced modulation of the primordial quantum fluctuations' amplitudes that we calculate was effectively fixed by measurement through environmental decoherence near horizon crossing.
The imprint in the CMB scales as $\sigma^\alpha$ with $\alpha =1$, and here, $\sigma = L_\mathrm{Planck}/L_\mathrm{Hubble} \sim H/\Omega$ could be as large as $10^{-5}$ for realistic values of $H$ and $\Omega$.
While still far from being measurable, it is conceivable that such an effect might eventually become observable, thereby providing some access to Planck-scale physics.   

\begin{figure}
\includegraphics[width=\columnwidth]{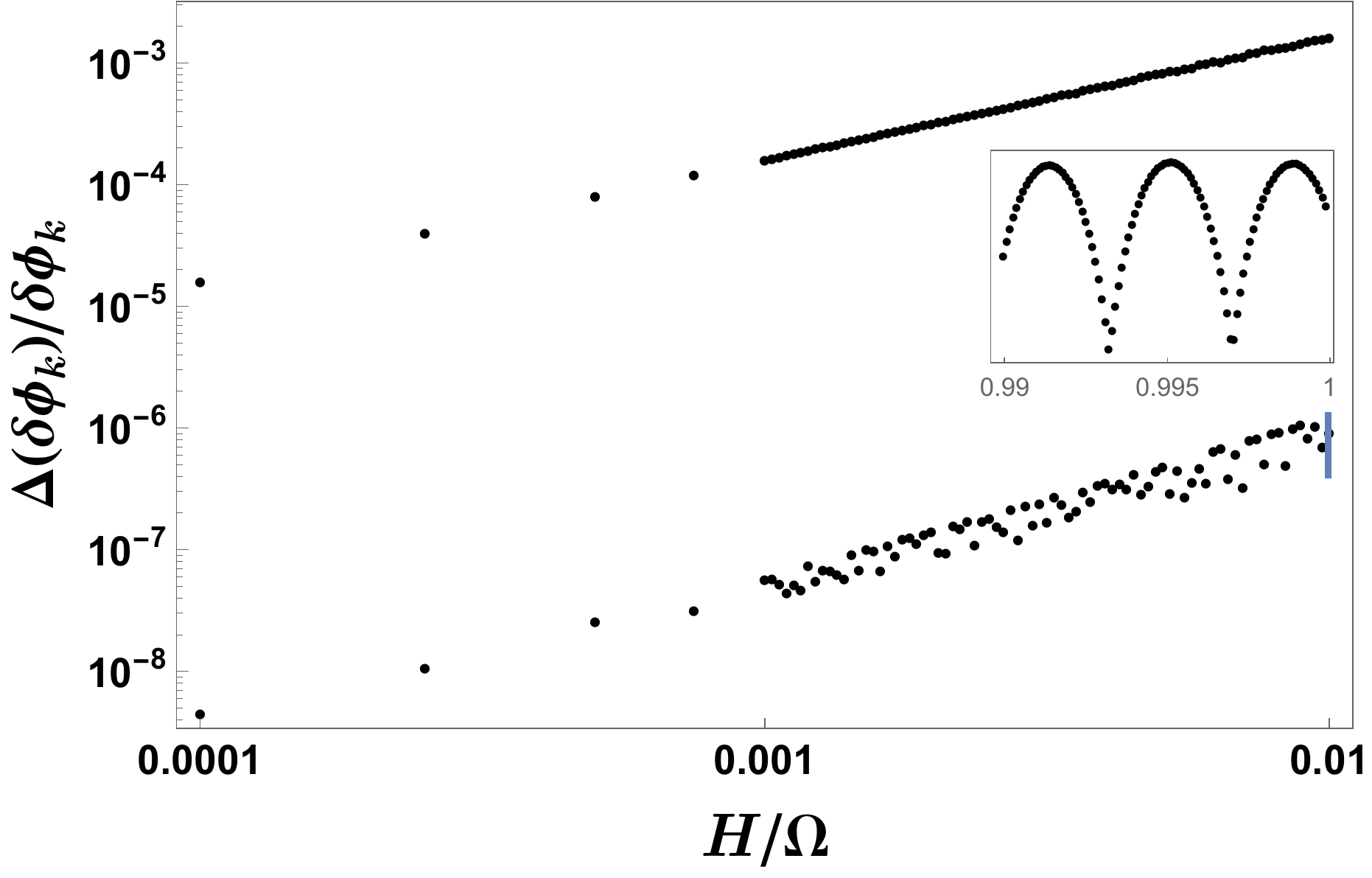}
\caption{Relative change in $\delta \phi_k$ as a function of the ratio of Planck to Hubble scales. The upper series of points is at horizon crossing ($k|\eta| = 1$), and the lower series is well past horizon crossing ($k|\eta| = 1/20$). The inset shows a sliver of the lower series at higher resolution, where the apparent scatter in the data is resolved as rapid oscillations.}
\label{fig:scaling}
\end{figure}

We determined the order, $\alpha$, by holding $k|\eta|$ fixed and plotting $\Delta(\delta\phi_k)/\delta\phi_k$ as a function of $H/\Omega$. This is shown in \Fig{fig:scaling}, where the scaling behavior can be read off:
The effect scales almost exactly linearly, i.e., $\alpha \approx 1$, when $k |\eta |$ is fixed close to horizon crossing.

\begin{figure}
\includegraphics[width=\columnwidth]{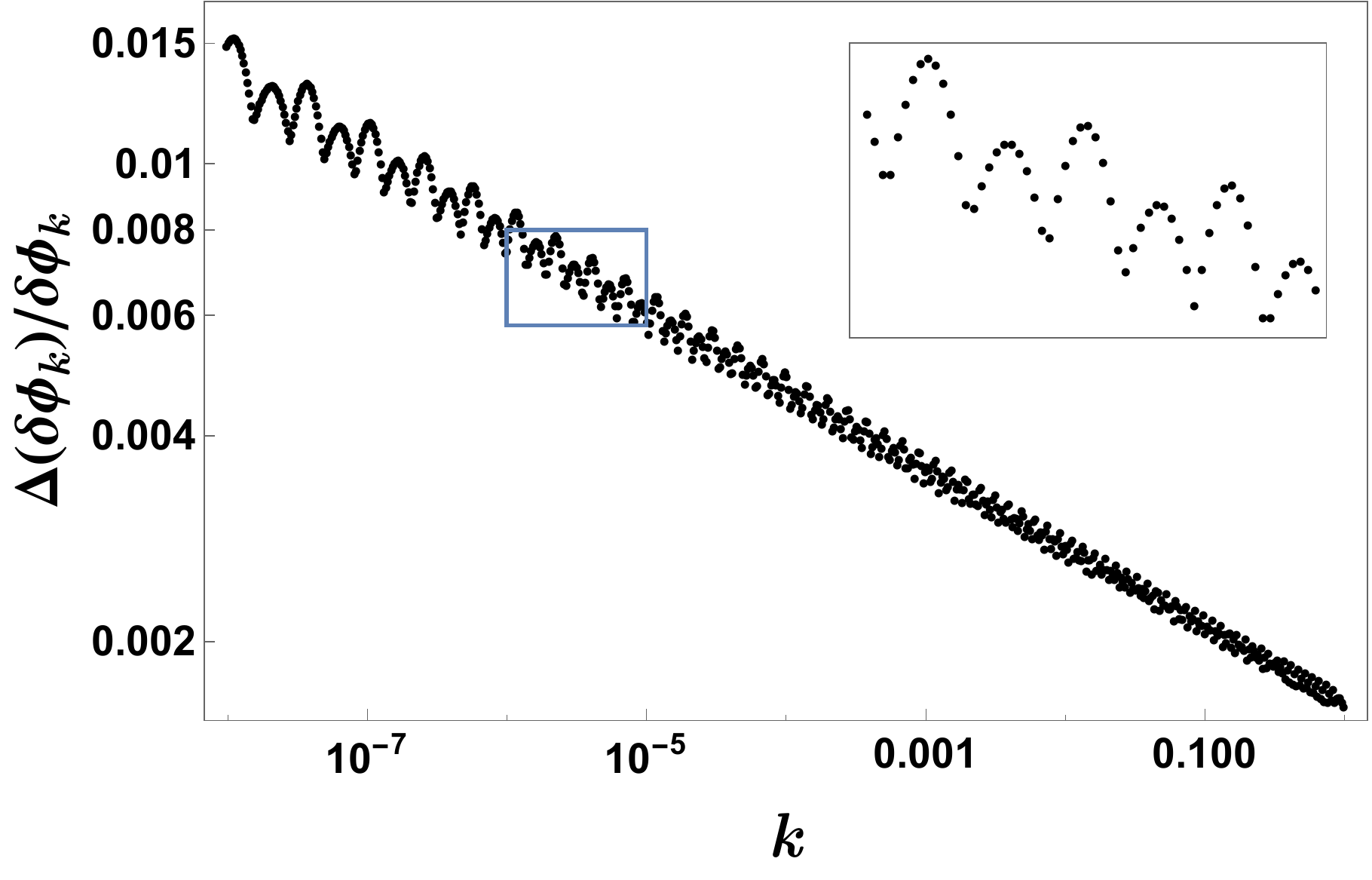}
\caption{Relative change in the fluctuation spectrum for a slowly-varying Hubble parameter. For each $k$, $|\eta|$ is set to $1/k$ and the Hubble parameter is matched to that of a power law spacetime: $H(\eta) = p[C(p-1)|\eta|]^{1/(p-1)}$, where $a(t) = Ct^p$ in cosmic time. Here we set $p=10$ and we fixed $C$ such that $H(-1)=1$, so as to coincide with horizon crossing in \Fig{fig:reldiff}. The value of $H$ ranges from 1 to 7.74 as $k$ goes from 1 to $10^{-8}$; this change is tracked by the linear trend in $\Delta(\delta \phi_k)/\delta \phi_k$. The inset shows a detail of the oscillations on top of the trend.}
\label{fig:adiabaticH}
\end{figure}

We now consider the realistic case of inflationary spacetimes with a slowly-varying Hubble parameter.
The exact calculations would be challenging since the corresponding d'Alembertian would be computationally even more difficult to diagonalize. We therefore model this case with an ``adiabatic'' approximation in which the spacetime is instantaneously de Sitter at every conformal time $\eta$, but in which we let $H$ slowly vary as a function of $\eta$.
\Fig{fig:adiabaticH} shows a plot of $\Delta(\delta \phi_k)/\delta \phi_k$ for this model.
At each $k$, $|\eta|$ is set to $1/k$ and the Hubble parameter is set to the value taken at the mode's horizon crossing by the time-varying Hubble parameter $H(\eta)$ of a power law spacetime.
As the magnitude of $\Delta(\delta \phi_k)/\delta \phi_k$ tracks the effective time-varying Hubble parameter, an interesting characteristic pattern of oscillations appears.
Intuitively, these oscillations may be thought of as arising from the time-varying number of Planckian wavelengths that fit into a Hubble length.
If observed, such oscillations in the CMB spectra may serve as an experimental signature of a covariant natural UV cutoff that could not easily be alternatively explained through a plausible inflaton potential.
Our results are consistent with prior literature \cite{Martin2001, Brandenberger2001, Brandenberger2002} in that we also predict superimposed oscillations. However, our new predictions for the type and magnitude of such oscillations are obtained covariantly and our predictions, including the prediction that the effect is of first order ($\alpha = 1$), are now free of potential artifacts due to covariance breaking.

\emph{Acknowledgements:}
This material is based upon work supported by the U.S. Department of Energy, Office of Science, Office of High Energy Physics, under Award Number DE-SC0011632, as well as by the Walter Burke Institute for Theoretical Physics at Caltech and the Gordon and Betty Moore Foundation through Grant 776 to the Caltech Moore Center for Theoretical Cosmology and Physics.
AK and ACD acknowledge support from the Natural Sciences and Engineering Research Council of Canada (NSERC). RTWM acknowledges support by the National Research Foundation (NRF) of South Africa, CPRR grant 90551.

\bibliographystyle{utphys.bst}
\bibliography{prl-refs}

\end{document}